\documentclass[useAMS,usenatbib]{mn2e}
\usepackage{amsmath,amssymb,graphicx,bm}
\usepackage{amsfonts}
\usepackage{subfigure}
\usepackage{multirow}
\usepackage{mathrsfs}

\title[The age problem in $\Lambda$CDM model]{The age problem in $\Lambda$CDM model}
\author[Yang and Zhang]{Rong-Jia Yang $^{1}$\thanks{E-mail:
yangrj08@gmail.com (RJY); zhangsn@tsinghua.edu.cn
(SNZ)}, Shuang Nan Zhang $^{2,\ 3}$\\
$^{1}$College of Physical Science and Technology, Hebei University, Baoding 071002, China\\
$^{2}$Key Laboratory for Particle Astrophysics, Institute of High
Energy Physics, Chinese Academy of Sciences,
Beijing 100049, China\\
$^{3}$Department of Physics, University of Alabama in Huntsville,
Optics Building 201C, Huntsville, AL 35899, USA}

\begin{document}

\date{Accepted date. Received  date}

\pagerange{\pageref{firstpage}--\pageref{lastpage}} \pubyear{2009}

\maketitle

\label{firstpage}

\begin{abstract}
The age problem in the $\Lambda$CDM model is re-examined. We define the elapsed time $T$ of an object is its age
plus the age of the Universe when it was born. Therefore in any cosmology, $T$ must be smaller than the age of
the Universe. For the old quasar APM~08279+5255 at $z=3.91$, previous studies have determined the best-fit value
of $T$, 1 $\sigma$ lower limit and the lowest limit to $T$ are 2.3, 2.0 and 1.7 Gyr, respectively. Constrained
from SNIa$+R+A+d$, SNIa$+R+A+d+H(z)$, and WMAP5+2dF+SNLS+HST+BBN, the $\Lambda$CDM model can only accommodate
$T(z=3.91)=1.7$ Gyr at 1 $\sigma$ deviation. Constrained from WMAP5 results only, the $\Lambda$CDM model can
only accommodate $T(z=3.91)=1.7$ Gyr at 2 $\sigma$ deviation. In all these cases, we found that $\Lambda$CDM
model accommodates the total age ($14$ Gyr for $z=0$) of the Universe estimated from old globular clusters, but
cannot accommodate statistically the 1 $\sigma$ lower limit to the best-fit age of APM~08279+5255 at $z=3.91$.
These results imply that the $\Lambda$CDM model may suffer from an age problem.
\end{abstract}


\section{Introduction}
Over the past decade, there are two most important reasons, one is
the ``age problem'', the other is the ``dark energy problem'', to
rule out with great confidence a large class of cold dark matter
(CDM) cosmological models. A matter-dominated spatially flat
Friedmann-Robertson-Walker (FRW) Universe (with age $T=2/3H_0$), for
example, is ruled out unless $h<0.48$, compared with the 14 Gyr age
of the Universe inferred from old globular clusters (Pont et al
1998). This is the ``age problem''. If one considers the age of the
Universe at high redshift, for instance, the 3.5 Gyr-old radio
galaxy 53W091 at $z=1.55$ and 4 Gyr-old radio galaxy 53W069 (Dunlop
et al 1996; Spinrad et al 1997), this problem becomes even more
acute.

The ``dark energy problem'' results from an increasing number of
independent cosmological observations, such as measurements of
intermediate and high redshift supernova Ia (SNIa), measurements of
the Cosmic Microwave Background (CMB) anisotropy, and the current
observations of the Large-Scale Structure (LSS) in the Universe.
These cosmological observations have consistently indicated that the
around $70\%$ of the present Universe energy content, with a
positive energy density but a negative pressure (called dark
energy), is homogeneously distributed in the Universe and is causing
the accelerated expansion of the Universe. The simplest and most
theoretically appealing candidate of dark energy is the vacuum
energy (or the cosmological constant $\Lambda$) with a constant
equation of state (EoS) parameter $w=-1$. This scenario is in
general agreement with the current astronomical observations, but
has difficulties to reconcile the small observational value of dark
energy density with estimates from quantum field theories (Peebles
and Ratra 2003; Carroll 2001; Padmanabhan 2003; Sahni and
Starobinsky 2000; Ishak 2007). The existence of such a ``dark
energy'' not only explains the accelerated expansion of the Universe
and the inflationary flatness prediction $\Omega_{\rm
total}\simeq1$, but also reconciles the ``age problem''. However,
the discovery of an old quasar, the APM~08279+5255 at $z=3.91$, has
once again led to an ``age problem''. The age of this quasar was
initially estimated to be around 2-3 Gyr (Komossa and Hasinger
2003); Friaca et al. (2005) re-evaluated the age of APM~08279+5255
to be 2.1 Gyr by using an improved method (we will discuss the
possible range of its age in more details later). For the currently
accepted values of the matter density parameter $\Omega_{\rm
m}=0.27\pm0.04$ (Spergel et al. 2003) and of the Hubble parameter
$H_0=72\pm8$ km s$^{-1}$Mpc$^{-1}$ (Freedman et al. 2001), most of
the existing dark energy scenarios cannot accommodate its age at
such a high redshift if imposing a prior on $H_0$, such as
$\Lambda$CDM model (Friaca et al. 2005; Alcaniz et al. 2003),
parameterized variable dark energy models (Dantas et al. 2007, Barboza and Alcaniz 2008), 
quintessence (Capozziello et al. 2007; Jesus et al.
2008), the $f(R)=\sqrt{R^2-R^2_0}$ model (Movahed et al. 2007),
braneworld models (Movahed and Ghassemi 2007; Pires et al. 2006;
Movahed and Sheykhi 2008; Alam and Sahni 2006), holographic dark
energy model (Wei and Zhang 2007), and other models (Sethi et al.
2005; Abreu et al. 2009; Santos et al. 2008).

But if one take values of $\Omega_{\rm m}$ and $H_0$ with $1\sigma$ deviation below $\Omega_{\rm m}=0.27\pm0.04$
(Spergel et al 2003) and $H_0=72\pm8$ km s$^{-1}$Mpc$^{-1}$ (Freedman et al. 2001), the age problem in
$\Lambda$CDM model (Friaca et al. 2005) or in holographic dark energy model can be alleviated (Wei and Zhang
2007). In other words, the ``age problem'', to a certain extent, is dependent on the values of the matter
density parameter and the Hubble constant one takes. Because the estimations of Hubble constant may have
somewhat large systematic errors at present, there are still debates on the value of $H_0$ in literatures. To
consider the ``age problem'' in a consistent way, unlike done previously by taking a set of cosmological
parameters a prior (Friaca et al. 2005; Wei and Zhang 2007), we do not take any special value of $\Omega_{\rm
m}$ or $H_0$ with prejudice. We instead  obtain directly observational constrains on $H_0$ and $\Omega_{\rm m}$
from SNIa, CMB, baryon acoustic oscillation (BAO), and $H(z)$ data points in the framework of the $\Lambda$CDM
model, then investigate the ``age problem'' in the parameter space allowed by these observations. We will also
discuss the ``age problem'' in $\Lambda$CDM model with the parameters from the five-year WMAP (WMAP5) data and
other observations.

The structure of this paper is as follows. In section II, we
consider constrains on the parameters of the $\Lambda$CDM model from
SNIa, CMB, BAO, and $H(z)$ observations, and present the parameters
from the five-year WMAP data. Using these best-fit values, the ``age
problem'' is discussed, and the possible range of the age of the
quasar APM~08279+5255 is also addressed in section III. Conclusions
and discussions are given in section IV.

\section{Observational constrains on $\Lambda$CDM}

In this section, we will consider observational constrains on
$\Lambda$CDM with SNIa, parameters measured from CMB and BAO, and
$H(z)$ data. We will also list the parameters constrained from WMAP5
data and other observations in the framework of the $\Lambda$CDM
model.

\subsection{Observational constrains on $\Lambda$CDM from SNIa, $R$, $A$, $d$, and $H(z)$}
To consider observational bounds on $\Lambda$CDM model for a flat
Universe, we use the recently published $182$ gold SNIa data with 23
SNIa at $z\gtrsim 1$ obtained by imposing constraints $A_{\rm
v}<0.5$ (excluding high extinction) (Riess et al. 2007). Each data
point at redshift $z_i$ includes the Hubble-parameter free distance
modulus $\mu_{\rm obs}(z_i)$ ($\equiv m_{\rm obs}-M$, where $M$ is
the absolute magnitude) and the corresponding error $\sigma^2(z_i)$.
The resulting theoretical distance modulus $\mu_{\rm th}(z)$ is
defined as
\begin{eqnarray}
\mu_{\rm th}(z)\equiv 5\log_{10}d_{\rm L}(z)+25,
\end{eqnarray}
where the luminosity distance in units of Mpc is expressed as
\begin{equation}
d_{\rm L}(z)=(1+z)\int^{z'}_0 \frac{dz'}{H},
\end{equation}
where $H=H_0E$ with $E=[\Omega_{\rm m}(1+z)^3+(1-\Omega_{\rm m})]$,
here $\Omega_{\rm m}$ includes baryon and cold dark matter. We treat
$H_0$ as a parameter and do not marginalize it over.

In order to break the degeneracies among the parameters, we consider
three $H_0$-independent parameters. One is the shift parameter $R$
measured from CMB observation, defined as (Bond et al. 1997;
Melchiorri and Griffiths 2001)
\begin{eqnarray}
R\equiv\Omega^{1/2}_{\rm m}\int^{z_{\rm r}}_0\frac{dz}{E(z)}.
\end{eqnarray}
where $z_{\rm r}=1089$ is the redshift of recombination. The shift parameter $R$ was found to be $R=1.70\pm0.03$
(Wang 2006) from WMAP three-year data recently. The other two $H_0$-independent parameters are $A$ parameter and
the distance ratio $d$, which are closely related to the measurements of the BAO peak in the distribution of
Sloan Digital Sky Survey (SDSS) luminous red galaxies (LRG), and are defined as respectively
\begin{eqnarray}
A&\equiv&\Omega^{1/2}_{\rm
m}(z^2_{1}E(z_1))^{-1/3}\left(\int^{z_1}_0dz/E(z)\right)^{2/3},\\
d&\equiv&(z_1/E(z_1))^{1/3}\frac{[\int^{z_1}_0dz/E(z)]^{2/3}}{\int^{z_{\rm
r}}_0dz/E(z)},
\end{eqnarray}
where $z_1=0.35$ is the effective redshift of the LRG sample. Measured from the SDSS BAO, the $A$ parameter and
the distance ratio $d$ were found to be $A=0.469\pm0.017$ and $d=0.0979\pm0.0036$ (Eisenstein et al 2005).

We also consider 9 $H(z)$ data points in the range $0\lesssim z\lesssim 1.8$ as shown in Table \ref{tab1}
(Jimenez et al. 2003; Simon et al. 2005; Abraham et al. 2004; Treu et al. 1999; Dunlop et al. 1996; Spinrad et
al. 1997; Nolan et al. 2003; Samushia and Ratra 2006).
\begin{table*}
\begin{tabular}{c|lllllllll}\hline\hline
 $z$ &\ 0.09 & 0.17 & 0.27 & 0.40 & 0.88 & 1.30 & 1.43
 & 1.53 & 1.75\\ \hline
 $H(z)$ &\ 69 & 83 & 70
 & 87 & 117 & 168 & 177 & 140 & 202\\ \hline
 1 $\sigma$ &$\pm 12$ & $\pm 8.3$ & $\pm 14$
 & $\pm 17.4$ & $\pm 23.4$ & $\pm 13.4$ & $\pm 14.2$
 & $\pm 14$ &  $\pm 40.4$\\ \hline\hline
\end{tabular}
\caption{\label{tab1} The observational $H(z)$ (km
s$^{-1}$Mpc$^{-1}$) data with 1 $\sigma$ uncertainty (Jimenez et al.
2003; Simon et al. 2005; Samushia and Ratra 2006)}
\end{table*}
These 9 $H(z)$ data points have been used to test dark energy models
recently.

These three $H_0$-independent parameters are extensively used to constrain dark energy models (e.g., Liddle et
al. 2006; Nesseris and Perivolaropoulos 2004; Yang et al. 2008). Some points regarding the use of these
parameters have been raised and discussed. This is an issue that deserves additional clarification. However,
many authors have shown that these parameters are effective to break the degeneracies among the parameters
(Liddle et al. 2006; Nesseris and Perivolaropoulos 2004; Yang et al. 2008).

Since the SNIa, CMB, BAO, and 9 $H(z)$ data points are effectively
independent measurements, we can simply minimize their total
$\chi^{2}$ value given by
\begin{eqnarray}
\label{21}\chi^2(\Omega_{\rm m}, H_0)=\chi^{2}_{\rm R}+\chi^{2}_{\rm
A}+\chi^{2}_{\rm d}+\chi^{2}_{\rm SNIa}+\chi^{2}_{\rm H},
\end{eqnarray}
to find the best-fit values of the parameters of the $\Lambda$CDM
model, where
\begin{eqnarray}
\chi^{2}_{\rm d}&=&\left(\frac{d-302.2}{1.2}\right)^2,\\
\chi^{2}_{\rm R}&=&\left(\frac{R-1.70}{0.03}\right)^2,\\
\chi^{2}_{\rm A}&=&\left(\frac{A-0.0979}{0.0036}\right)^2,\\
\chi^{2}_{\rm H}&=&\sum^{N}_{i=1}\frac{[H_{\rm obs}(z_i)-H_{\rm
th}(z_i)]^2}{\sigma^2_{{\rm H}_i}}
\end{eqnarray}
and
\begin{eqnarray}
\chi^{2}_{\rm SNIa}&=&\sum^{N}_{i=1}\frac{(\mu^{\rm obs}_{\rm L}(z_i)-\mu^{\rm th}_{\rm L}(z_i))^2}
{\sigma^2_i}.
\end{eqnarray}

Fitting SNIa, CMB, and BAO, we find the best-fit values of the parameters at 68.3\% confidence: $\Omega_{\rm
m}=0.288\pm0.008$ and $H_{\rm 0}=63.7\pm3$ km s$^{-1}$Mpc$^{-1}$ with $\chi^2_{\rm min}=163.39$ ($\chi^2_{\rm
min}/$dof=0.89, $p(\chi^2>\chi^2_{\rm min})=0.87$), as shown in Table \ref{tab2}.

\begin{table*}
\begin{tabular}{c|c|c|c|c}\hline\hline
 Observations     &$\Omega_{\rm m}$   & $H_{0}$   &$\chi^2_{\rm min}/$dof   & $p$ \\ \hline
 SNIa$+R+A+d$ &$0.288\pm0.008$  &$63.7\pm3$  &0.89  &0.85 \\ \hline
 SNIa$+R+A+d+H(z)$ &$0.302\pm0.009$  &$63.6\pm3$  &$0.95$  &$0.73$ \\ \hline\hline
\end{tabular}
\caption{\label{tab2}The best values of the parameters ($\Omega_{\rm
m}$, $H_0$) of $\Lambda$CDM model with the corresponding
$\chi^2_{\rm min}/$dof and p($\chi^2>\chi^2_{\rm min}$) fitting from
SNIa$+R+A+d$ and SNIa$+R+A+d+H(z)$ observations with 1 $\sigma$
confidence level, here $H_0$ with dimension km s$^{-1}$Mpc$^{-1}$.}
\end{table*}

If the 9 $H(z)$ data points are also included in fitting, we find the best-fit values of the parameters at
68.3\% confidence: $\Omega_{\rm m}=0.302\pm0.009$ and $H_{\rm 0}=63.6\pm3$ km s$^{-1}$Mpc$^{-1}$ with
$\chi^2_{\rm min}=181.57$ ($\chi^2_{\rm min}/$dof=0.95, $p(\chi^2>\chi^2_{\rm min})=0.73$), as shown in Table
\ref{tab2}.

All these results are consistent with $\Omega_{\rm m}=0.27\pm0.04$ (Spergel et al. 2003) measured from WMAP in
the $\Lambda$CDM model and $H_0=62.3\pm 1.3$ (random )$\pm5.0$ (systematic) km s$^{-1}$Mpc$^{-1}$ from HST
Cepheid-calibrated luminosity of Type Ia SNIa observations (Sandage et al. 2006).

\subsection{Observational constrains on $\Lambda$CDM from WMAP5 data and other observations}
WMAP data are analyzed in the framework of the $\Lambda$CDM model,
$H_0$ (or $h$) and $\Omega_{\rm m}h^2$ can be constrained directly.
Here we quote the results obtained by Dunkley et al. (2008). They
constrained baryon and cold dark matter density parameters
($\Omega_{\rm b}h^2$ and $\Omega_{\rm c}h^2$), dimensionless Hubble
parameter ($h$), cosmological constant density parameter
($\Omega_{\Lambda}$), the scalar spectral index ($n_s$), the optical
depth to reionization ($\tau$), and the linear theory amplitude of
matter fluctuations on $8h^{-1}$ Mpc scales ($\sigma_8$), with
five-year WMAP (WMAP5) data. They found the best-fit value of
$\Omega_{\rm b}h^2$, $\Omega_{\rm c}h^2, h$ as $\Omega_{\rm
b}h^2=0.02273\pm0.00062$, $\Omega_{\rm c}h^2=0.1099\pm0.0062$,
$h=0.719^{+0.026}_{-0.027}$.

Recently, the case of coupling neutrino mass was considered to
constrain cosmological parameters. For example, Vacca et al. (2009)
constrained $\Omega_{\rm b}h^2, \Omega_{\rm c}h^2, \tau, n_s,$, the
ratio of the sound horizon to the angular diameter distance at
recombination ($\theta$), the amplitude of the scalar fluctuations
at a scale of $\kappa=0.002$ Mpc$^{-1}$ ($A_S$), the sum of neutrino
mass ($M_{\nu}$), the energy scale in dark energy (cosmological
constant) potentials ($\Lambda$), and the coupling parameter between
CDM and dark energy ($\beta$), with observations of WMAP5, 2dF
galaxy redshift survey (2dF), Supernova Legacy Survey (SNLS), Hubble
Space Telescope (HST), and Big Bang Nucleosynthesis (BBN). They
found the best-fit values of $\Omega_{\rm b}h^2$, $\Omega_{\rm
c}h^2$, $H$ as $10^2\Omega_{\rm b}h^2=2.258\pm0.061$, $\Omega_{\rm
c}h^2=0.1098\pm0.0040$, $H_0=70.1\pm2.1$ km s$^{-1}$Mpc$^{-1}$. We
will use these results to discuss the ``age problem" in $\Lambda$CDM
in the next section.

\section{Age problem in $\Lambda$CDM}
Old high-redshit objects are usually used to test dark energy model or constrain parameters (see e. g. Lima et al. 2009).
Recently, the quasar APM~08279+5255 at $z=3.91$ have been used to
test many dark energy models, such as $\Lambda$CDM model (Friaca et
al. 2005; Alcaniz et al. 2003), $\Lambda(t)$ model (Cunha and Santos 2004), parametrized variable Dark Energy
Models (Barboza and Alcaniz 2008; Dantas 2007), quintessence
(Capozziello et al. 2007; Jesus et al. 2008), the
$f(R)=\sqrt{R^2-R^2_0}$ model (Movahed et al. 2007), braneworld
modes (Movahed and Ghassemi 2007; Pires et al. 2006; Movahed and
Sheykhi 2008; Alam and Sahni 2006), holographic dark energy model
(Wei and Zhang 2007, Granda et al. 2009), and other models (Sethi et
al. 2005; Abreu et al. 2009; Santos et al. 2008).
It was shown that the quasar APM~08279+5255 ($z=3.91$) cannot
accommodated most dark energy models. In order to understand the
problem better, we first discuss the possible range of the age of
APM~08279+5255.

\subsection{The age of APM~08279+5255}
APM~08279+5255 is an exceptionally luminous broad absorption line (BAL) quasar at redshift $z=3.91$. From
\emph{XMM-Newton} observations of APM~08279+5255, Hasinger et al. (2002) have derived an iron overabundance of
Fe/O of $3.3\pm0.9$ (here the abundance ratio has been normalized to the solar value) for the BAL system. Using
an Fe/O=3 abundance ratio, derived from X-ray observations, Komossa and Hasinger (2003) estimated the age of the
quasar APM~08279+5255 to lie within the interval 2-3 Gyr. An age of 3 Gyr is inferred from the temporal
evolution of Fe/O ratio in the giant elliptical model (M4a) of Hamann and Ferland (1993, hereafter HF93). For
the `extreme model' M6a of HF93, the Fe/O evolution would be faster, and Fe/O=3 is already reached after 2 Gyr.

Friaca et al. (2005) re-evaluated the age of APM~08279+5255 by using
a chemodynamical model for the evolution of spheroids. An age of 2.1
Gyr is set by the condition that Fe/O abundance ratio of the model
reaches 3.3, which is the best-fitting value obtained in Hasinger et
al. (2002). An age of 1.8 Gyr is set when the Fe/O abundance ratio
reaches 2.4, 1 $\sigma$ deviation from the best-fitting value 3.3.

An age of 1.5 Gyr is set when the Fe/O abundance ratio reaches 2.
But there is a correlation between Fe/O values and the neutral
hydrogen column density $N_{\rm H}$, in the sense that lower Fe/O
values are obtained for higher values of $N_{\rm H}$. Because of
this, a value of Fe/O as low as two is highly improbable, as it
would require $N_{\rm H}$ in excess of $1.2\times10^{22}$ cm$^{-2}$
(see from Fig.~3 of Hasinger et al. 2002), which seems to be ruled
out from the determinations of $N_{\rm H}=(5.3-9.1)\times10^{22}$
cm$^{-2}$ by other \emph{Chandra} and \emph{XMM} observations. Even
considering only the \emph{XMM}2 data set, the lowest value of Fe/O
is $2.4$ at $1.28\times10^{22}$ cm$^{-2}$ for 1 $\sigma$ deviation
from the best-fit 3.3 (see Fig.~3 of Hasinger et al. 2002).

Based on the above discussions, we obtain the following estimates of
{\it the age of APM~08279+5255 since the initial star formation and
stellar evolution in the galaxy}: (1) the best estimated value is
2.1 Gyr; (2) 1 $\sigma$ lower limit is 1.8 Gyr; (3) the lowest limit
is 1.5 Gyr.

When using the best available WMAP5 polarization results (Dunkley et
al. 2008), the polarization optical depth $\tau$ values imply a peak
epoch of reionizing photons at $z=10.8 \pm 1.4$. However considering
that the much smaller and more preliminary WMAP first year data set
implied  $z=17 \pm 10$ for the peak epoch of reionization, we take
the peak reionization redshift as between $z=8-14$. Suppose that the
Population III stars are mostly responsible for the reionization,
then we can further estimate that these star formation processes
start as early as $z=15-17$ in high density peaks. This also agrees
with recent results based on the new Hubble WFC3/IR imaging in the
Ultradeep Field, which suggest that HST has now in fact seen the
tail-end of this reionizing population at $z=8-10$ (Yan et al.
astro-ph/0910.0077). We therefore conclude that the quasar
APM~08279+5255 started its initial formation process at least
0.2-0.3 Gyr since the beginning of the Universe.

Therefore, we define the elapsed time $T$ of an object is its age
plus the age of the universe when it was born. Since the spirit of
testing cosmological models with ages of astrophysical objects is to
examine if such objects can have sufficient time to be formed given
the age of the Universe at that specific redshift,  and we need to
calculate the elapsed time $T$ of the objects since the beginning of
the Universe, and compare this time ($T$ in all following figures)
with the elapsed time (age) of the Universe at that specific
redshift since the beginning of the Universe. Based upon the above
discussions, we obtain the following estimates of {\it the elapsed
time $T$ of APM~08279+5255 since the beginning of the Universe}: (1)
the best estimated value is 2.3 Gyr; (2) 1 $\sigma$ lower limit is
2.0 Gyr; (3) the lowest limit is 1.7 Gyr.

\subsection{Parameters constrained from SNIa, $R$, $A$, $d$, and $H(z)$}
In this subsection, we use APM~08279+5255 to discuss the age problem
in the $\Lambda$CDM model in the parameter space ($\Omega_{\rm
m}-H_0$ plane) allowed by SNIa, CMB, BAO, and $H(z)$ observations,
obtained in the previous section.

The age-redshift relation for a spatially flat, homogeneous, and isotropic universe with the vacuum energy reads
\begin{eqnarray}
T(z)=\int^{\infty}_z\frac{dz'}{H_0(1+z')\sqrt{\Omega_{\rm m}(1+z')^3+(1-\Omega_{\rm m})}}.
\end{eqnarray}
With this equation, one can calculate the age of the Universe at any
redshift in the framework of the $\Lambda$CDM model.

\begin{figure}
\includegraphics[width=8.5cm]{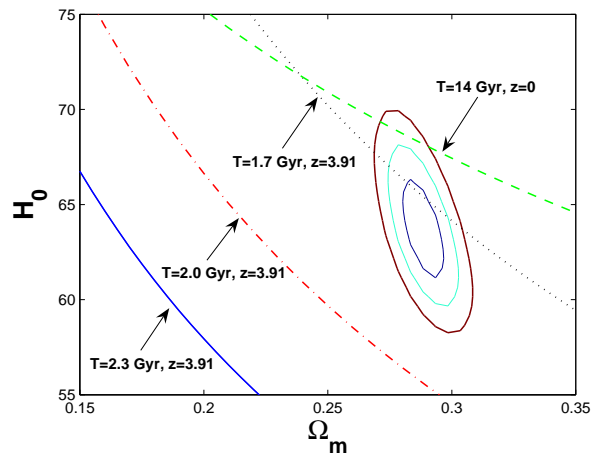}
\caption{The $68.3\%$, $95.4\%$ and $99.7\%$ confidence regions in the $\Omega_{\rm m}$-$H_{0}$ (km
s$^{-1}$Mpc$^{-1}$) plane fitting from SNIa, CMB,and BAO observations. The dot, dash-dot, and solid line
represent the cosmological parameter values corresponding to the ages of 1.7, 2.0 and 2.3 Gyr of the Universe at
$z=3.91$, respectively. The dash line represents 14.0 Gyr at $z=0$. \label{Fig3}}
\end{figure}

Taking $\Omega_{\rm m}=0.288$ and $H_{\rm 0}=63.7$ km
s$^{-1}$Mpc$^{-1}$ obtained from fitting SNIa, CMB, and BAO
observations, we find the present age of the Universe is $T=15.0$
Gyr, larger than 14 Gyr estimated from old globular clusters (Pont
et al. 1998), but $\Lambda$CDM just accommodates the lowest limit to
the elapsed time ($T=1.7$ Gyr) of APM~08279+5255 at the 1 $\sigma$
deviation, as shown in figure \ref{Fig3}. There is no way with these
parameters $\Lambda$CDM can accommodate statistically even the 1
$\sigma$ lower limit to the elapsed time: $T=2.0$ Gyr.

\begin{figure}
\includegraphics[width=8.5cm]{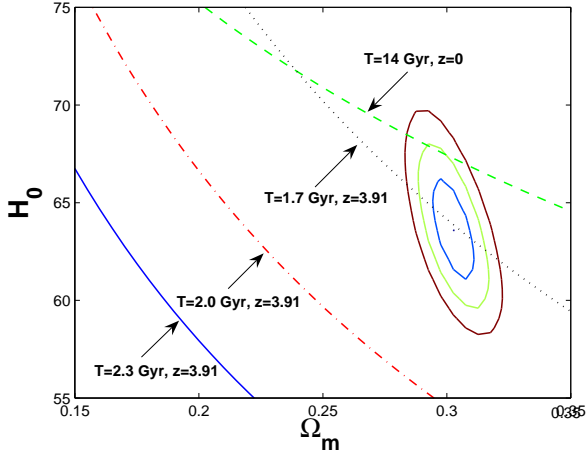}
\caption{The $68.3\%$, $95.4\%$ and $99.7\%$ confidence regions in the $\Omega_{\rm m}$-$H_{0}$ (km
s$^{-1}$Mpc$^{-1}$) plane fitting from SNIa, CMB, BAO, and $H(z)$ observations, compared with the same lines as
in Fig 1. \label{Fig4}}
\end{figure}

Similarly, taking $\Omega_{\rm m}=0.302$ and $H_{\rm 0}=63.6$ km
s$^{-1}$Mpc$^{-1}$ obtained from fitting SNIa, CMB, BAO, and $H(z)$
observations, we find the present age of the Universe is $T=14.8$
Gyr, again larger than 14 Gyr estimated from old globular cluster
(Pont et al. 1998), but $\Lambda$CDM just accommodates the lowest
limit to the elapsed time ($T=1.7$ Gyr) of APM~08279+5255 at the 1
$\sigma$ deviation, as shown in figure \ref{Fig4}. There is also no
way with these parameters $\Lambda$CDM can accommodate statistically
even the 1 $\sigma$ lower limit to the elapsed time: $T=2.0$ Gyr.

As shown in figures \ref{Fig3} and \ref{Fig4}, the only way to
reconcile the elapsed time $T$ of APM~08279+5255 with the age of the
Universe at $z=3.91$ in the $\Lambda$CDM model, is to take smaller
values of $H_0$ and $\Omega_{\rm m}$, which certainly contradict
many other independent observations. We therefore conclude that the
$\Lambda$CDM model suffers possibly from an age problem. 
These discussions are summarized in Table \ref{tab3}.

\subsection{Parameters constrained from WMAP5+2dF+SNLS+HST+BBN}
Here we use APM~08279+5255 to discuss the age problem in the
$\Lambda$CDM model with the parameters constrained from WMAP5 data
and other observations. Taking the best-fit values of $\Omega_{\rm
m}h^2=0.1326$ and $h=0.719$, obtained from fitting WMAP5 data
(Dunkley et al. 2008), we find the present age of the Universe is
$T=13.7$ Gyr, which is less than 14 Gyr estimated from old globular
clusters (Pont et al. 1998). However with the 1 $\sigma$ deviations
of $\Omega_{\rm m}h^2$ and $h$, $\Lambda$CDM can accommodate the age
of old globular clusters; for instance, taking $\Omega_{\rm
m}h^2=0.12643$ and $h=0.692$, we obtain the present age of the
Universe is $T=14.1$ Gyr. This situation is depicted in upper right
corner of Fig.~3. However, with such 1 $\sigma$ deviations,
$\Lambda$CDM still cannot accommodate even the lowest limit to the
age of APM~08279+5255 at redshift $z=3.91$; only with 2 $\sigma$
deviations, $\Lambda$CDM can accommodate the lowest limit: $T=1.7$
Gyr. Even with 2 $\sigma$ deviations of $\Omega_{\rm m}h^2$ and $h$,
$\Lambda$CDM cannot accommodate the 1 $\sigma$ lower limit to the
best-fit age: $T=2.0$ Gyr; statistically there is no feasibility to
be consistent with the best-fit value: $T=2.3$ Gyr. Therefore the
age of APM~08279+5255 conflicts with WMAP5 results severely within
the $\Lambda$CDM model.

\begin{figure}
\includegraphics[width=8.5cm]{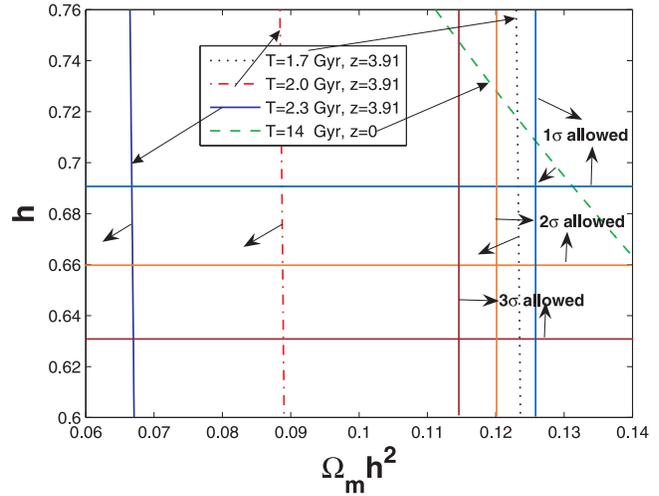}
\caption{The upper right corner shows the 1 $\sigma$ allowed
parameter space with $\Omega_{\rm m}h^2\geq0.1264$ and $h\geq0.692$
from WMAP5, which can accommodate $T=14.0$ Gyr at $z=0$. However
$T\geq1.7$ Gyr at $z=3.91$ for APM~08279+5255 cannot be accommodated
statistically. \label{Fig5}}
\end{figure}

Similarly, taking the best-fit values of $\Omega_{\rm m}h^2=0.11208$
and $H_0=70.1$ km s$^{-1}$Mpc$^{-1}$, obtained from fitting
WMAP5+2dF+SNLS+HST+BBN (Vacca et al. 2009), we find the present age
of the Universe is $T=14.5$ Gyr, larger than 14 Gyr estimated from
old globular clusters (Pont et al. 1998). With 1 $\sigma$ deviations
of $\Omega_{\rm m}h^2$ and $h$, $\Lambda$CDM can accommodate the
lowest limit: $T=1.7$ Gyr, as shown in the upper right corner of
Fig.~4. For $T=2.0$ and 2.3 Gyr, the situation is qualitatively
similar to that shown in Fig.~3, with however slightly less severe
conflicts with the $\Lambda$CDM model. These discussions are summarized in Table \ref{tab3}.

\begin{figure}
\includegraphics[width=8.5cm]{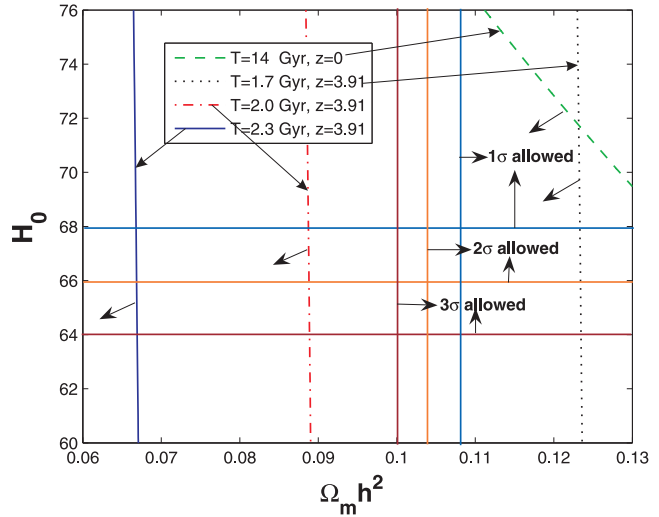}
\caption{The upper right corner shows the 1 $\sigma$ allowed parameter space with $\Omega_{\rm m}h^2\geq0.1081$
and $H_0\geq68$ (km s$^{-1}$Mpc$^{-1}$) from WMAP5+2dF+SNLS+HST+BBN, which can accommodate both $T=14.0$ Gyr at
$z=0$ and $T=1.7$ Gyr at $z=3.91$ for APM~08279+5255. However $T\geq2.0$ Gyr at $z=3.91$ cannot be accommodated
statistically. \label{Fig6}}
\end{figure}

\begin{table*}
\begin{tabular}{|c|cc|c|c|ccc|}
\hline Data & Cosmological parameters & & Age of old globular &
\multicolumn{3}{|c|}{Elapsed time of APM~08279+5255 (Gyr)}
\\&&& clusters 14 Gyr & Best value & 1 $\sigma$ lower limit & the lowest
limit
\\&&& & $T=2.3$ & $T=2.0$ & $T=1.7$\\
\hline SNIa$+R+A+d$ & $\Omega_{\rm m}=0.288\pm0.008$ & 1 $\sigma$ &
yes & no & no & yes
\\& $H_{\rm 0}=63.7\pm3$ & 2 $\sigma$ &   yes & no &  no & yes
\\&&  3 $\sigma$ &  yes & no & no & yes
\\\hline
SNIa$+R+A+d+H(z)$ & $\Omega_{\rm m}=0.302\pm0.009$ & 1 $\sigma$ &
yes & no & no & yes
\\& $H_{\rm 0}=63.6\pm3$ & 2 $\sigma$ &  yes & no & no &yes
\\&& 3 $\sigma$ & yes & no & no & yes
\\\hline
WMAP5 & $\Omega_{\rm b}h^2=0.02273\pm0.00062$ &  1 $\sigma$ &  yes &
no & no & no
\\& $\Omega_{\rm c}h^2=0.1099\pm0.0062$ & 2 $\sigma$ & yes &  no & no & yes
\\& $h=0.719^{+0.026}_{-0.027}$ & 3 $\sigma$ &  yes & no & no &  yes \\
\hline WMAP5+SNLS+ & $10^2\Omega_{\rm b}h^2=2.258\pm0.061$ & 1
$\sigma$ &  yes & no & no & yes
\\HST+BBN+2dF & $\Omega_{\rm c}h^2=0.1098\pm0.0040$ & 2 $\sigma$ &  yes &  no & no & yes
\\& $H_0=70.1\pm2.1$ & 3 $\sigma$ & yes & no & no & yes \\
\hline
\end{tabular}
\caption{\label{tab3} Summary on the constraints to the ages of the Universe at $z=0$ and $z=3.91$ in the
$\Lambda$CDM model, using different combinations of observational data sets. Here $H_0$ takes the dimension of
km s$^{-1}$Mpc$^{-1}$.}
\end{table*}

\section{conclusions and discussions}
As previous many works have shown, the age problem in dark energy
models is dependent on the values of Hubble constant and matter
density one taken, at least to a certain degree. Because the
estimations of Hubble parameter may have somewhat large systematic
errors currently, there are still debates on the value of $H_0$ in
literatures. In the paper, we re-examine the age problem in the
$\Lambda$CDM model in a consistent way, i.e., without requiring {\it
a prior} values of the parameters. We define the elapsed time $T$ of
an object is its age plus the age of the universe when it was born.
Therefore in any cosmology, $T$ must be smaller than the age of the
Universe. For the old quasar APM~08279+5255 at $z=3.91$, previous
studies have determined the best-fit value, 1 $\sigma$ lower limit
and the lowest limit to $T$ are 2.3, 2.0 and 1.7 Gyr, respectively.
In Table 1 we summarize the constraints to the ages of the Universe
at $z=0$ and $z=3.91$, using different combination of observational
data sets.

Fitting SNIa, CMB, and BAO observations, we have obtained the
best-fit values of the parameter at 68.3\% confidence: $\Omega_{\rm
m}=0.288\pm0.008$ and $H_{\rm 0}=63.7\pm3$ km s$^{-1}$Mpc$^{-1}$
with $\chi^2_{\rm min}=163.39$ ($p(\chi^2>\chi^2_{\rm min})=0.87$).
In the $\Omega_{\rm m}-H_{\rm 0}$ parameter space allowed by these
observations, the $\Lambda$CDM model accommodates the total age (14
Gyr for $z=0$) of the Universe estimated from old globular clusters
(Pont et al. 1998), but just accommodate the lowest limit to the
elapsed time ($T=1.7$ Gyr) of APM~08279+5255 at 1 $\sigma$
deviation. There is no way, with these parameters, $\Lambda$CDM can
accommodate statistically even the 1 $\sigma$ lower limit to the
elapsed time: $T=2.0$ Gyr.

If $H(z)$ observations are also included, the best-fit values of the
parameter at 68.3\% confidence are: $\Omega_{\rm m}=0.302\pm0.009$
and $H_{\rm 0}=63.6\pm3$ km s$^{-1}$Mpc$^{-1}$ with $\chi^2_{\rm
min}=181.57$ ($p(\chi^2>\chi^2_{\rm min})=0.73$). In this case the,
the $\Lambda$CDM model is also consistent with the total age of the
Universe estimated from old globular clusters (Pont et al. 1998),
but just accommodate the lowest limit to the elapsed time ($T=1.7$
Gyr) of APM~08279+5255 at 1 $\sigma$ deviation.

Constrained from WMAP5 only, the $\Lambda$CDM model can accommodate
the total age of the Universe estimated from old globular clusters
but cannot accommodates the lowest limit to the age ($T=1.7$ Gyr) of
APM~08279+5255 at redshift $z=3.91$ at 1 $\sigma$ deviation; only 2
$\sigma$ deviations can accommodate even the absolute lower limit of
$T=1.7$ Gyr. Even 2 $\sigma$ deviations of $\Omega_{\rm m}h^2$ and
$h$ cannot accommodate the 1 $\sigma$ lower limit of $T=2.0$ Gyr.

Constrained from WMAP5+2dF+SNLS+HST+BBN, the $\Lambda$CDM model can
accommodate the total age of the Universe estimated from old
globular clusters (Pont et al. 1998), and can accommodate the lowest
limit to the elapsed time ($T=1.7$ Gyr) of APM~08279+5255 at
redshift $z=3.91$ at 1 $\sigma$ deviation. The situation is
qualitatively similar to the case constrained from WMAP5 only, with
however slightly less severe conflicts with the $\Lambda$CDM model.

The only way to reconcile the elapsed time $T$ of APM~08279+5255 with the age of the Universe at $z=3.91$ in the
$\Lambda$CDM model, is to take very small values of $H_0$ and $\Omega_{\rm m}$, which certainly contradict many
other independent observations. We therefore conclude that the $\Lambda$CDM model suffers from a problem with
the estimated age of APM~08279+5255 at redshift $z=3.91$, based on the currently best available data for the
Hubble constant $H_0$ and the matter density $\Omega_{\rm m}$ (recently Riess et al. (2009) obtained
$H_0=74.2\pm3.6$ km s$^{-1}$Mpc$^{-1}$, which may lead to more serious age problem in $\Lambda$CDM model). These
results can be tested with future cosmological observations. Of course, new and more reliable determination of
the age of APM 08279+5255 are also needed. We mention in passing that when the Dunlop et al. (1996) paper first
came out, the 3.5 Gyr age of 53W091 at $z=1.5$ was a significant problem for a non-$\Lambda$ cosmology; however
soon thereafter papers came out that did spectral energy distribution (SED) fitting of its Keck spectrum that allowed new age estimates of
around 1.8 Gyr (Yi et al. 2000, Bruzual and Magris 1999).

\section*{Acknowledgments}
We thank Y. Liu for help discussions. The anonymous referees are
thanked for his/her insightful and constructive criticisms and
suggestions. This study is supported in part by Directional Research
Project of the Chinese Academy of Sciences under project No.
KJCX2-YW-T03, by the National Natural Science Foundation of China
under grant Nos. 10821061, 10733010, 10725313, by 973 Program of
China under grant 2009CB824800, by Research Fund for Doctoral
Programs of Hebei University No. 2009-155, and by Open Research
Topics Fund of Key Laboratory of Particle Astrophysics, Institute of
High Energy Physics, Chinese Academy of Sciences, No.
0529410T41-200901.

\bsp
\label{lastpage}

\end{document}